\documentclass[twocolumn]{jpsj2}
\usepackage{graphicx}
\usepackage{amsmath}
\def\runauthor {%
Y. {\sc Asano}, et. al. }

\title
{%
Josephson effect in quasi one-dimensional unconventional superconductors }

\author{%
Yasuhiro {\sc Asano}$^{1}$
Yukio {\sc Tanaka}$^{2}$,
Yasunari {\sc Tanuma}$^{2}$,
Kazuhiko {\sc Kuroki}$^{4}$
 and
Hiroki {\sc Tsuchiura}$^{5}$ 
}

\inst
{
$^{1}$ Department of Applied Physics, Hokkaido University, 
Sapporo 060-8628, Japan.
\\
$^{2}$ Department of Applied Physics, Nagoya 
University, Nagoya, 464-8603, Japan and Crest, Japan Science and Technology Corporation (JST), 
Nagoya 464-8063 Japan.
\\
$^{3}$ Institute of Physics, Kanagawa University, Rokkakubashi, Yokohama, 221-8686, Japan. 
\\
$^{4}$ Department of Applied Physics and Chemistry, The university of Electro-Communications, 
Chofu 182-8568, Japan.
\\
$^{5}$ S.I.S.S.A. - Via Beirut 2-4 34014 Trieste, Italy.
}
\recdate
{\today}

\abst{%
Josephson effect in junctions of quasi one-dimensional triangular lattice superconductors
is discussed, where the theoretical model corresponds to organic superconductors 
(TMTSF)$_2$PF$_6$. We assume the quarter-filling electron band and $p$, $d$ and $f$ wave 
like pairing symmetries in organic superconductors.
To realize the electronic structures in organic superconductors, we introduce the asymmetric
hopping integral, ($t'$) among second nearest lattice sites.
At $t'=0$, the Josephson current in the $d$ wave symmetry saturates in low temperatures,
whereas those in the $p$ and the $f$ wave symmetries show the low-temperature anomaly due to 
the zero-energy state at the junction interfaces.
The low-temperature anomaly appears even in the $d$ wave symmetry in the presence
of $t'$, whereas the anomaly is suppressed in the $f$ wave symmetry.
The shape of the Fermi surface is an important factor for the formation
of the ZES in the quarter-filling electron systems. 
 }
\kword{% 
organic superconductor, zero-energy states, low-temperature anomaly, quarter-filling
}
\begin{document}
%\sloppy
\maketitle

\section{Introduction}

Quasi one-dimensional (Q1D) superconductors (TMTSF)$_2$X (X=PF$_6$, ClO$_4$, etc.)~\cite{jerome,bechgaad}.have recently attracted much attention as a possible
spin-triplet superconductor.
A spin-triplet pairing has been suggested from an observation of large $H_{c2}$~\cite{ijlee1}
and unchanged Knight shift across $T_c$~\cite{ijlee2}.
Unconventional pairing with nodes on the Fermi surface has been suggested from an NMR 
measurement~\cite{takigawa}, while a thermal conductivity measurement has reported the absence of nodes on the Fermi surface~\cite{belin}. In addition, a recent experiment showed the 
zero-bias conductance peak (ZBCP) in tunnel spectra~\cite{ha}. 
Theoretically, a $p$ wave paring in which nodes of 
the pair potential can be made to avoid intersecting the Q1D Fermi surface has been 
proposed in an early stage~\cite{abrikosov,hasegawa,lebed}.
On the other hand, a spin-singlet $d$ wave pairing mediated by spin fluctuations has been proposed 
by several authors~\cite{shimahara,kuroki3,kino}.
This is because superconductivity lies right next to the $2k_F$ spin density wave (SDW) 
phase in the pressure-temperature phase diagram. Moreover, one of the present authors has recently 
proposed that~\cite{kuroki4}
 a triplet $f$ wave pairing may dominate over the $d$ and the $p$ wave in (TMTSF)$_2$PF$_6$ due 
to a combination of a Q1D Fermi surface, coexistence of the $2k_F$ SDW
and the $2k_F$ charge density wave (CDW) suggested from a diffuse X-ray 
scattering~\cite{pouget,kagoshima}, and an anisotropy in the spin fluctuations.
Thus the situation is not settled either experimentally or theoretically.
In recent papers, we proposed the tunneling conductance experiment to determine which one of 
pairing symmetries is realized in (TMTSF)$_2$X~\cite{tanuma1,tanuma2} because
the tunnel spectra are sensitive to the internal degree of freedom of Cooper pairs.

A great variety in the Andreev reflection~\cite{andreev} 
is one of interesting features in superconductors with unconventional
pairing symmetries. 
The pair potential of a transmitted quasiparticle in the electron branch differs 
from that in the hole branch because of unconventional pairing symmetries.
When the two pair potentials have opposite signs to each other, 
constructive interference effects of a quasiparticle near the surface of 
superconductors lead to the zero-energy state 
(ZES)~\cite{buchholtz,hu,tanaka0,rpp,lofwander,ya03-4}.
The ZES is observed as 
the ZBCP in tunneling conductance spectra.
Since the high-$T_c$ superconductors have the $d$ wave pairing 
symmetry~\cite{tsuei,wollman,harlingen,barone,sigrist,tanaka1994},
the ZBCP was found in a number of experiments~\cite{rpp,lofwander,kashiwaya,kashi96,kashi95,alff,wang,wei,iguchi,geerk,mao,Ekin,Sawa1,Sawa2,Aubin}.
So far a considerable number of theoretical studies have been made on the ZES 
itself~\cite{buchholtz,hu,ya03-4} and 
related phenomena of transport properties in both 
spin-singlet~\cite{matsumoto,nagato,ohhashi,sign1,sign3,zhu1,stefanakis,wu,dong,barash4,higashi,t1,kusakabe,honerkamp,sengupta,shirai,tsuchiura95} and 
spin-triplet~\cite{yama1,yama2,yama3,barash3,ya02-2,ya03-1,ya03-3} 
unconventional superconductor junctions.
Effects of ferromagnets attaching to
superconductors~\cite{zhu,kashi2,zutic,y1,y2,y4,h1,h2}, those of broken time-reversal symmetry 
states (BTRSS)~\cite{fogelstrom,covington,biswas,dagan,sharoni,kohen,matsumoto2,laughlin,lubimova,kitaura,TJ1,TJ2,Tanuma2001}, those of magnetic fields~\cite{qazilbash,alff2,sawa3,YT022} and those of the random 
potentials
~\cite{barash2,golubov,poenicke,yamada,tanaka01b,luck,asai00,ya02-1,ya03-2,ya01-2,ya02-3,circuit}
on the ZBCP are hot topics in recent studies. 
Since the ZES appears just on the Fermi energy, 
it drastically affects transport properties through the interface of 
unconventional superconductor junctions. 
The low-temperature anomaly of the Josephson current between the two unconventional 
superconductors is explained in terms of the resonant tunneling of Cooper pairs through the 
ZES~\cite{barash,tanaka1,tanaka2,tanaka3,tanaka4,ya01-2,ya02-2,ya01-3}.
The Josephson effect in high-$T_c$ superconductors has been reported in a number of
papers~\cite{riedel,samanta,stefana2,Ilichev1,Ilichev2,Arie,hilgenkamp1,tafuri,smilde,hilgenkamp2,imaizumi}.

In theories, the Andreev reflection in unconventional superconductors has been studied
based on the free electron model in which the quadratic dispersion relation and the 
isotropic Fermi surface are assumed. Although the Fermi surface in real unconventional
superconductors are not isotropic at al, 
the free electron model well explains characteristic behaviors of the tunnel conductance 
and the Josephson current. This may be because such transport properties are 
sensitive to the pairing symmetries of superconductors. However recent studies 
show several exceptions~\cite{tanuma2,shirai}. 
For instance in high-$T_c$ superconductor junctions, the free electron 
model does not explain the Josephson current calculated
on the two-dimensional tight-binding model for some cases~\cite{shirai}.
The low-temperature anomaly of the Josephson current is expected in the free
electron model when the $a$ axis of the high-$T_c$ superconductors 
oriented from the junction interface normal.
The anomaly is washed out by the Friedel oscillations of the wave function 
in the lattice simulation for specific orientation angles.
In addition, the density of states at the surface of the superconductors 
is sensitive to the shape of the Fermi surface in Q1D organic superconductors
(TMTSF)$_2$X when triangular lattice structures are taken into account.
In such situation, the Josephson current is also expected to be sensitive to
the pairing symmetries and the shape of the Fermi surface.
In this paper, we discuss the direct-current Josephson effect motivated by
the surface density of states in Q1D superconductors~\cite{tanuma1,tanuma2,vaccarella}.
So far, a Josephson effect in organic superconductors has been reported in the theoretical
paper~\cite{kwon}. The effects of electronic structures in QID lattice, however, are not
taken into account. 

This paper is organized as follows. In Sec.~2, we described the Josephson
junctions of organic superconductors by the Bogoliubov-de Gennes equation
on the two-dimensional lattice. The Josephson current is discussed for
$p$, $d$ and $f$ wave symmetries in Sec,~3. In Sec.~4, we discuss the calculated
results. We summarize this paper in Sec.~5.

\section{Model and method}
\begin{figure}[htb]
\begin{center}
\includegraphics[width=7.0cm]{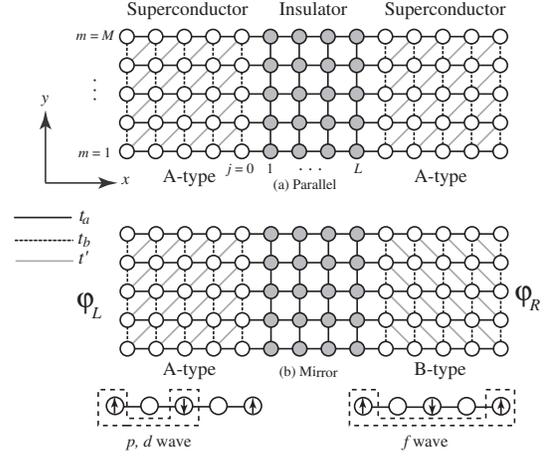}
\end{center}
\caption{ A schematic figure of the SIS junction of the organic
superconductor.
}
\label{fig:system}
\end{figure}

Let us consider Q1D superconductor / insulator / superconductor
(SIS) junctions as shown in 
Fig.~\ref{fig:system}, where $\boldsymbol{r}=j \bar{\boldsymbol{x}}+ m\bar{\boldsymbol{y}}$
labels a lattic site, where $\bar{\boldsymbol{x}}$ and $\bar{\boldsymbol{y}}$ are unit vectors
in the $x$ and the $y$ directions, respectively.
The two superconductors (i.e., $-\infty \leq j \leq 0$ 
and $L+1 \leq j \leq \infty$) are separated by
the insulator (i.e., $1 \leq j \leq L$). We assume the periodic boundary condition in the $y$ 
direction and the number of lattice sites in the $y$ direction is $M$. 
The junctions are described by the mean-field Hamiltonian
\begin{align}
H_{\textrm{BCS}}=& \frac{1}{2}\sum_{\boldsymbol{r},\boldsymbol{r}'} 
 \left[ \tilde{c}_{\boldsymbol{r}}^\dagger\;  {h}_{\boldsymbol{r},\boldsymbol{r}'} 
 \; \hat{\sigma}_0
  \; \tilde{c}_{\boldsymbol{r}'}^{ }  -  
\tilde{c}_{\boldsymbol{r}}^{t}\;  {h}^\ast_{\boldsymbol{r},\boldsymbol{r}'}  \; \hat{\sigma}_0
 \; \left\{ \tilde{c}_{\boldsymbol{r}'}^\dagger \right\}^{t} \right] \nonumber \\
 + \frac{1}{2} \sum_{\boldsymbol{r},\boldsymbol{r}' \in \textrm{S}}&
 \left[ \tilde{c}_{\boldsymbol{r}}^\dagger 
\hat{\Delta}_{\boldsymbol{r},\boldsymbol{r}'}
\left\{\tilde{c}_{\boldsymbol{r}'}^\dagger\right\}^{t}
- \left\{\tilde{c}_{\boldsymbol{r}}\right\}^{t} 
\hat{\Delta}^\ast_{\boldsymbol{r},\boldsymbol{r}'}
\tilde{c}_{\boldsymbol{r}'} \right], \label{bcs}\\
{h}_{\boldsymbol{r},\boldsymbol{r}'}=& -t_{\boldsymbol{r},\boldsymbol{r}'}
+ (\epsilon_{\boldsymbol{r}}- \mu_{\boldsymbol{r}})\delta_{\boldsymbol{r},\boldsymbol{r}'},\\
\hat{\Delta}_{\boldsymbol{r},\boldsymbol{r}'}=& 
\left\{ \begin{array}{ccc}
i \boldsymbol{d}_{\boldsymbol{r},\boldsymbol{r}'}
\cdot \hat{\boldsymbol{\sigma}} \hat{\sigma}_2 &:& \textrm{triplet} \\
i d_{\boldsymbol{r},\boldsymbol{r}'}\hat{\sigma}_2 &:& \textrm{singlet},
\end{array}\right. \\
\tilde{c}_{\boldsymbol{r}}=&\left( \begin{array}{c} c_{\boldsymbol{r},\uparrow} \\
c_{\boldsymbol{r},\downarrow}\end{array}\right),
\end{align}
where $c_{\boldsymbol{r},\sigma}^{\dagger}$ ($c_{\boldsymbol{r},\sigma}^{ }$) 
is the creation (annihilation) operator of an electron at $\boldsymbol{r}$ with 
spin $\sigma =$ ( $\uparrow$ or $\downarrow$), $\hat{\sigma}_0$ is the 
$2\times 2$ unit matrix representing the spin space, $\hat{\sigma}_j$ with $j=1 \sim 3$ are 
the Pauli matrices and S in the summation denotes the superconductors.
The Fermi energy in the superconductor is $\mu_{\boldsymbol{r}}=\mu_S$ for $j\leq 0$ and 
$j\geq L+1$, and that in the insulator is $\mu_{\boldsymbol{r}}=\mu_N$ for $1\leq j\leq L$.
In superconductors, we assume $\epsilon_{\boldsymbol{r}}=0$ and $t_a$ and $t_b$ are the hopping 
integral in the $x$ and $y$ directions, respectively. 
We also introduce asymmetric hopping integral ($t'$) between the second
nearest neighbors as shown in Fig.~\ref{fig:system} to realize electronic 
structures in (TMTSF)$_2$PF$_6$.
In the insulator, $\epsilon_{\boldsymbol{r}}=V_B$ for $1 \leq j\leq L$ denotes 
the barrier potential and the hopping integrals in the two directions are equal to $t_a$.
The pair potential in $p$, $d$ and $f$ wave symmetries are defined by,
\begin{align}
\boldsymbol{d}_{\boldsymbol{r},\boldsymbol{r}'}^{(p)}
=& \frac{\Delta}{2}\; e^{i\varphi_i}\; \textrm{sgn}(j-j')\; \delta_{|j-j'|,2}\; \delta_{m,m'} 
\; \boldsymbol{e}_3, \label{pairp}\\
d_{\boldsymbol{r},\boldsymbol{r}'}^{(d)}
=& \frac{\Delta}{2}\; e^{i\varphi_i}\; \delta_{|j-j'|,2}\; \delta_{m,m'}, \label{paird}\\
\boldsymbol{d}_{\boldsymbol{r},\boldsymbol{r}'}^{(f)}
=& \frac{\Delta}{2}\; e^{i\varphi_i}\; \textrm{sgn}(j-j')\; \delta_{|j-j'|,4}\; \delta_{m,m'}
\; \boldsymbol{e}_2, \label{pairf}
\end{align}
where $\varphi_i=\varphi_L$ or $\varphi_R$ is the macroscopic phase of 
superconductors and $\boldsymbol{e}_j$ with $j=1,$ 2 and 3 are unit vectors in the spin space.
 A schematic picture of Cooper pairs are shown in Fig.~\ref{fig:system}.
The Hamiltonian is diagonalized by the Bogoliubov transformation,
\begin{align}
\left[ 
\begin{array}{c}
\tilde{c}_{\boldsymbol{r}} \\
\left\{\tilde{c}^\dagger_{\boldsymbol{r}}\right\}^{t} 
\end{array}
\right]=& \sum_\lambda
\left[
\begin{array}{cc}
\hat{u}_\lambda(\boldsymbol{r}) & \hat{v}_\lambda^\ast(\boldsymbol{r}) \\
\hat{v}_\lambda(\boldsymbol{r}) & \hat{u}_\lambda^\ast(\boldsymbol{r}) 
\end{array}\right]
\left[ \begin{array}{c}
\tilde{\gamma}_{\lambda} \\
\left\{\tilde{\gamma}^\dagger_{\lambda}\right\}^{t} 
\end{array}
\right], \label{bt}\\
\tilde{\gamma}_{\lambda}=&\left( \begin{array}{c} \gamma_{\lambda,\uparrow} \\
\gamma_{\lambda,\downarrow}\end{array}\right),
\end{align}
where ${\gamma}^\dagger_{\lambda,\sigma}$ 
(${\gamma}_{\lambda,\sigma}$) is creation (annihilation) operator of a
Bogoliubov quasiparticle. In Eq.~(\ref{bt}),
$\hat{u}_\lambda$ and $\hat{v}_\lambda$ are the wavefunction of a quasiparticle which 
satisfy the Bogoliubov-de Gennes (BdG) equation~\cite{degennes}.
\begin{align}
\sum_{\boldsymbol{r}'}&
\left[ \begin{array}{cc}
{h}_{\boldsymbol{r},\boldsymbol{r}'}\;\hat{\sigma}_0 & 
\hat{\Delta}_{\boldsymbol{r},\boldsymbol{r}'}\\
-\hat{\Delta}^\ast_{\boldsymbol{r},\boldsymbol{r}'} & -{h}^\ast_{
\boldsymbol{r},\boldsymbol{r}'}\;\hat{\sigma}_0 \end{array}
\right]  
\left[ \begin{array}{c}
\hat{u}_\lambda(\boldsymbol{r}')  \\
\hat{v}_\lambda(\boldsymbol{r}')  
\end{array}\right] 
= E_\lambda \left[ \begin{array}{c}
\hat{u}_\lambda(\boldsymbol{r})  \\
\hat{v}_\lambda(\boldsymbol{r})  
\end{array}\right]. \label{bdg}
\end{align}
The eigenvalue $E_\lambda$ is independent of spin channels because we consider unitary
states in superconductors.
In what follows, we briefly discuss the method to calculate the Josephson current 
for the $d$ wave symmetry. The application to the $p$ and the $f$ wave symmetries is 
straightforward.
In the case of the $d$ wave symmetry, the BdG equation in Eq.~(\ref{bdg}) is decoupled to
two equations,
\begin{align}
\sum_{\boldsymbol{r}'}&
\left[ \begin{array}{cc}
{h}_{\boldsymbol{r},\boldsymbol{r}'} & d^{(d)}_{\boldsymbol{r},\boldsymbol{r}'}\\
\left(d^{(d)}_{\boldsymbol{r},\boldsymbol{r}'}\right)^\ast & -{h}^\ast_{
\boldsymbol{r},\boldsymbol{r}'}\end{array}
\right]  
\left[ \begin{array}{c}
\left(u_{11}\right)_\lambda(\boldsymbol{r}')  \\
\left(v_{21}\right)_\lambda(\boldsymbol{r}')  
\end{array}\right] \nonumber \\
=& E_\lambda \left[ \begin{array}{c}
\left(u_{11}\right)_\lambda(\boldsymbol{r})  \\
\left(v_{21}\right)_\lambda(\boldsymbol{r})  
\end{array}\right], \label{bdg-d1}
\end{align}
where $\left(u_{ij}\right)$, for example, represents an 
element of $\hat{u}$ in Eq.~(\ref{bt}) and 
$\left[ u_{21}, v_{11} \right]^{t}$ obeys essentially the same equation.
In the following, we suppress $(1,1)$ of $\left(u_{11}\right)$ and $(2,1)$ of 
$\left(v_{21}\right)$.
In this way, the wave function 
\begin{equation}
\boldsymbol{\Psi}_\lambda(j) = \left(\begin{array}{c}
u_\lambda(j\bar{\boldsymbol{x}}+1\bar{\boldsymbol{y}}) \\
\vdots\\
u_\lambda(j\bar{\boldsymbol{x}}+M\bar{\boldsymbol{y}}) \\
v_\lambda(j\bar{\boldsymbol{x}}+1\bar{\boldsymbol{y}}) \\
\vdots\\
v_\lambda(j\bar{\boldsymbol{x}}+M\bar{\boldsymbol{y}}) \\
\end{array}
\right),
\end{equation}
satisfies the BdG equation. For $j < -2$, for instance,
the BdG equation reads
\begin{align}
&\frac{\Delta}{2}\left(\begin{array}{cc}
\hat{0} & e^{i\varphi_L} \hat{1}\\ 
e^{-i\varphi_L} \hat{1} & \hat{0}
\end{array}\right)
\boldsymbol{\Psi}_\lambda(j+2) \nonumber\\
&+ \left(\begin{array}{cc}
\hat{T}_N(-) & \hat{0} \\
\hat{0} & -\hat{T}_N(-)
\end{array}\right) 
\boldsymbol{\Psi}_\lambda(j+1) \nonumber \\
&+\left(\begin{array}{cc}
-E_\lambda \hat{1} + \hat{E}_S & \hat{0} \\
\hat{0} & -E_\lambda \hat{1}-\hat{E}_S
\end{array}\right)
\boldsymbol{\Psi}_\lambda(j) \nonumber \\
&+\left(\begin{array}{cc}
\hat{T}_N(+) & \hat{0} \\
\hat{0} & -\hat{T}_N(+)
\end{array}\right) 
\boldsymbol{\Psi}_\lambda(j-1) \nonumber \\
&+\frac{\Delta}{2}\left(\begin{array}{cc}
\hat{0} & e^{i\varphi_L} \hat{1} \\
e^{-i\varphi_L} \hat{1} & \hat{0}
\end{array}\right)
\boldsymbol{\Psi}_\lambda(j-2)=0, 
\end{align}
\begin{align}
&\hat{T}_N(+) =
\left(\begin{array}{ccccc}
-t_a & -t' &  0 & \cdots & 0 \\
0    & -t_a & -t' & \cdots & 0\\
\vdots& \ddots&  \ddots & \ddots & \vdots \\
0& \cdots & 0 & -t_a & -t'\\
-t'& 0& \cdots & 0  &-t_a
\end{array}\right),\\ 
&\hat{T}_N(-) =
\left(\begin{array}{ccccc}
-t_a & 0 & \cdots & 0 & -t' \\
-t'& -t_a & 0 & \cdots & 0\\
\vdots& \ddots& \ddots & \ddots & \vdots \\
0& \cdots & -t' & -t_a & 0\\
0 & \cdots & 0& -t'  &-t_a
\end{array}\right),\\ 
&\hat{E}_S =
\left(\begin{array}{ccccc}
-\mu_S & -t_b & 0 & \cdots & -t_b \\
-t_b& -\mu_S & -t_b & \cdots & 0\\
\vdots& \ddots & \ddots & \ddots & \vdots \\
0 &\cdots & -t_b  &-\mu_S & -t_b\\
-t_b & 0& \cdots & -t_b  &-\mu_S
\end{array}\right),
\end{align}
where $\hat{1}$ and $\hat{0}$ are the $M\times M$ unit matrix and the zero matrix, 
respectively. 
To solve the BdG equation, we apply the recursive Green function 
method~\cite{lee,furusaki1,ya01-1} 
and calculate the Matsubara Green function in a matrix form
\begin{align}
\check{G}_{\omega_n}(j,j') 
=& \sum_\lambda \boldsymbol{\Psi}_\lambda(j)[i \omega_n - E_\lambda]^{-1}
\boldsymbol{\Psi}^\dagger_\lambda(j')
,\label{defg}
\end{align}
where $\omega_n=(2n+1)\pi T$ is the Matsubara frequency and $T$ is a temperature.
 Throughout this paper, we use 
the units of $\hbar=k_B=1$, where $k_B$ is the Boltzmann constant. 
The Josephson current in the insulator $(1<j<L)$, is 
given by~\cite{furusaki1,ya01-1,ya02-4}
\begin{equation}
J(j) = -ie T
 \sum_{\omega_n}t_a{\rm Tr}
\left[  \check{G}_{\omega_n}(j+1,j)- \check{G}_{\omega_n}(j,j+1)
\right]. \label{jq}
\end{equation}
We note that $J(j)$ is independent of $j$ when we consider the 
direct-current Josephson effect. 

\section{Josephson current}
The low-temperature anomaly of the Josephson current is a typical phenomenon in the
quantum transport between two unconventional superconductors~\cite{barash,tanaka1,tanaka2,tanaka3,tanaka4,ya01-2,ya02-2,ya01-3}. In this section, we discuss effects of the asymmetric 
second nearest neighbor hopping ($t'$) on the ZES at the interface and on the Josephson current. 
For finite $t'$, it is possible to consider two types of SIS junctions as shown in 
Fig.~\ref{fig:system} (a) and (b). The parallel junction consists of two A-type superconductors,
whereas the mirror-type junction consists of a A-type and a B-type superconductors. 
Generally speaking, the Josephson effect in the two junctions are not identical to 
each other.  
  
In what follows,
we choose parameters as $t_b=0.1t_a$, $\mu_S=-1.4099t_a$, $\mu_N=-2.0t_a$, 
$M=20$, $L=4$, and $V_B=2.0t_a$.
The electron density is fixed at the quarter-filling. The amplitude of the pair potential at the
zero temperature is $\Delta_0=0.1t_a$ and the dependence of the pair potential on temperatures 
is described by the BCS theory.

\subsection{$p$ wave symmetry}
\begin{figure}[htb]
\begin{center}
\includegraphics[width=7.0cm]{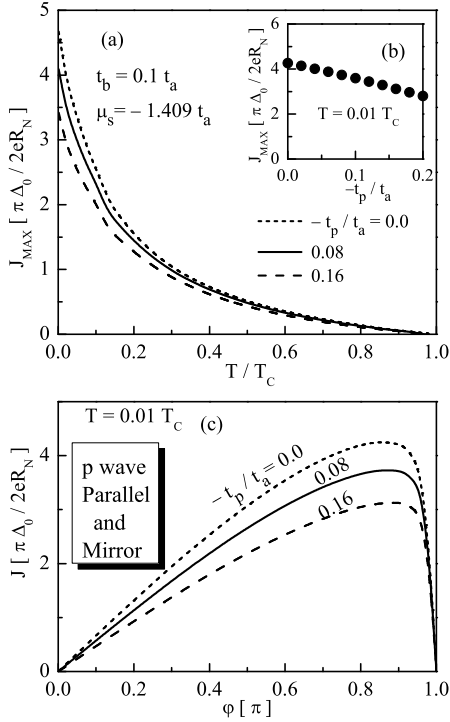}
\end{center}
\vspace{-0.5cm}
\caption{ Josephson current for the $p$ wave symmetry is shown, 
where $t_b=0.1t_a$, $\mu_S=-1.4099t_a$, $M=20$, $L=4$, and $V_B=2t_a$.
In (a), the maximum amplitude of the Josephson current ($J_{max}$) is plotted as a function of 
temperatures.
In (b), $J_{max}$ at $T=0.01T_c$ are shown as a function of $-t'/t_a$. 
The current-phase relation is calculated at $T=0.01T_c$ in (c).
}
\label{fig:pj}
\end{figure}

In Fig.~\ref{fig:pj}, we show the Josephson current
in the parallel junctions with the $p$ wave symmetry. 
\begin{figure}[htb]
\begin{center}
\includegraphics[width=8.0cm]{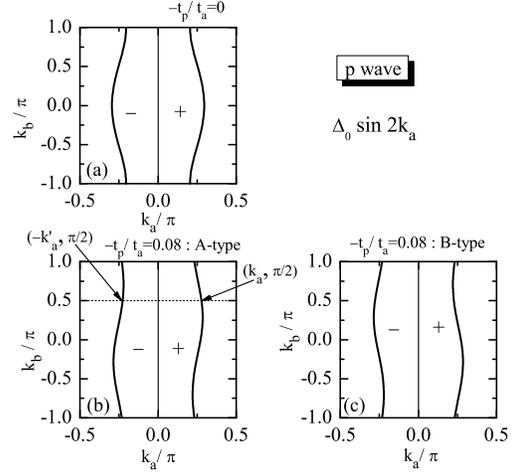}
\end{center}
\vspace{-0.5cm}
\caption{   
The Fermi surface of TMTSF for $t'=0$ is shown in (a). 
Those for $t'=-0.08t_a$ in the A-type and the B-type superconductors are 
shown in (b) and (c),
respectively. The pair potential in the $p$ wave symmetry is given by $\Delta \sin 2k_a$
which changes the sign at $k_a=0$ as indicated by + and - in the figures.
}
\label{fig:pfermi}
\end{figure}
The vertical axis is normalized by $\pi \Delta_0 / 2eR_N$, where $R_N$ is the 
normal resistance of junctions. We note in this vertical scale that the Josephson current of the 
$s$ wave symmetry with the isotropic Fermi surface is close to unity in the limit of
the zero temperature~\cite{ambegaokar}.
In real materials, the second nearest neighbor hopping should be $-t'/t_a$ = 0.08.
We also show results for $-t'/t_a$ = 0 and 0.16 for comparison.
In Fig.~\ref{fig:pj} (a), the maximum value of the Josephson current $(J_{max})$ 
is plotted as a function 
of temperatures, where $(J_{max})$ is estimated from the Josephson current as a function of
$\varphi=\varphi_L-\varphi_R$. The Josephson current increases 
with decreasing temperature irrespective of $t'$ and does not saturate even in low
temperatures. This behavior is called as the low-temperature
anomaly of the Josephson current and is owing to the ZES forming at the junction interface.
When the $\boldsymbol{d}$ vector has only one component in the free electron model,
a condition for appearance of the ZES is given by
\begin{equation}
\boldsymbol{d}(k_a,k_b) \cdot \boldsymbol{d}(-k_a,k_b) < 0, \label{zes0}
\end{equation}
where $\boldsymbol{d}(k_a,k_b)$ is the Fourier component of 
$\boldsymbol{d}_{\boldsymbol{r}-\boldsymbol{r}'}$, and $k_a$ and $k_b$ are the wavenumber
in the $x$ and $y$ directions, respectively. The two pair potentials in Eq.~(\ref{zes0})
corresponds to the pair potentials in the electron and the hole branch of a quasiparticle.
Since the translational invariance in the $y$ holds, $k_b$ is conserved in the transmission
and the reflection of a quasiparticle at the junction interfaces.
In the $p$ wave symmetry, the pair potential in Eq.~(\ref{pairp}) results in
\begin{equation}
\boldsymbol{d}^{(p)}(k_a,k_b) = \Delta \sin( 2k_a) \boldsymbol{e}_3. \label{pairp2}
\end{equation}
Since Eq.~(\ref{pairp2}) is an odd function of $k_a$, it satisfies Eq.~(\ref{zes0}) 
for all the 
Fermi surface as shown in 
Fig.~\ref{fig:pfermi}, where we draw the Fermi surface for $t'=0$ in (a),
$t'=-0.08t_a$ in the A-type superconductor in (b), and $t'=-0.08t_a$ in the B-type 
superconductor in (c) with the solid line. 
The pair potential in the $p$ wave symmetry has a node line at $k_a=0$ in both the 
free electron model and the lattice model. In the lattice model, however, 
Eq.~(\ref{pairp2}) has {\em additional node lines} at $k_a=\pm 0.5\pi$ because
the pairing interaction works between two electrons on the second nearest
neighbor sites at the quarter-filling as shown in Fig.~\ref{fig:system}.
We define the {\em additional node lines} as the node lines appear in the pair 
potentials because of the quarter-filled electron band on the Q1D lattice. 

The asymmetric hopping $(t')$, modulates the shape of the Fermi surface
as shown in Fig.~\ref{fig:pfermi} (b) and (c).
When the Fermi surface looses a symmetry with respect to $k_a=0$ as in (b) and (c), 
the condition for the ZES should be rewrote as
\begin{equation}
\boldsymbol{d}(k_a,k_b) \cdot \boldsymbol{d}(-k'_a,k_b) < 0, \label{zes1}
\end{equation}
where $k_a$ and $-k_a'$ are the wave numbers of the Fermi surface for fixed $k_b$.
In (b), the two wavenumbers are indicated by arrows on the Fermi surface for
$k_b=\pi/2$.
In the following, we call Eq.~(\ref{zes1}) as the zero-energy condition (ZEC).
Even in the presence of $t'$, the ZEC is always
satisfied because the {\em additional node lines} are far from the Fermi surface.
The amplitude of the Josephson current at $T=0.01T_c$ decreases with increasing $-t'/t_a$
as show in Fig.~\ref{fig:pj} (b). This is mainly because $R_N$ decreases with increasing
$-t'/t_a$. In our calculation, $R_N$ in units of $h/e^2$ are 1.04, 0.83 and 0.63 for
$-t'/t_a=$ 0, 0.08 and 0.16, respectively.
In Fig.~\ref{fig:pj} (c), the current-phase relation is calculated at $T=0.01T_c$.
The current-phase relation in a low temperature deviates from the sinusoidal function 
because the resonant tunneling through the ZES enhances the multiple Andreev reflection 
between the two superconductors. We note that the results in the mirror-type junction are
identical to those in Fig.~\ref{fig:pj} (a)-(c). In the $p$ wave symmetry, the Josephson current
are not qualitatively changed by introducing $t'$ because the ZES at the interface governs 
the characteristic behavior of the Josephson current.

\subsection{$d$ wave symmetry}
\begin{figure}
\begin{center}
\includegraphics[width=7.0cm]{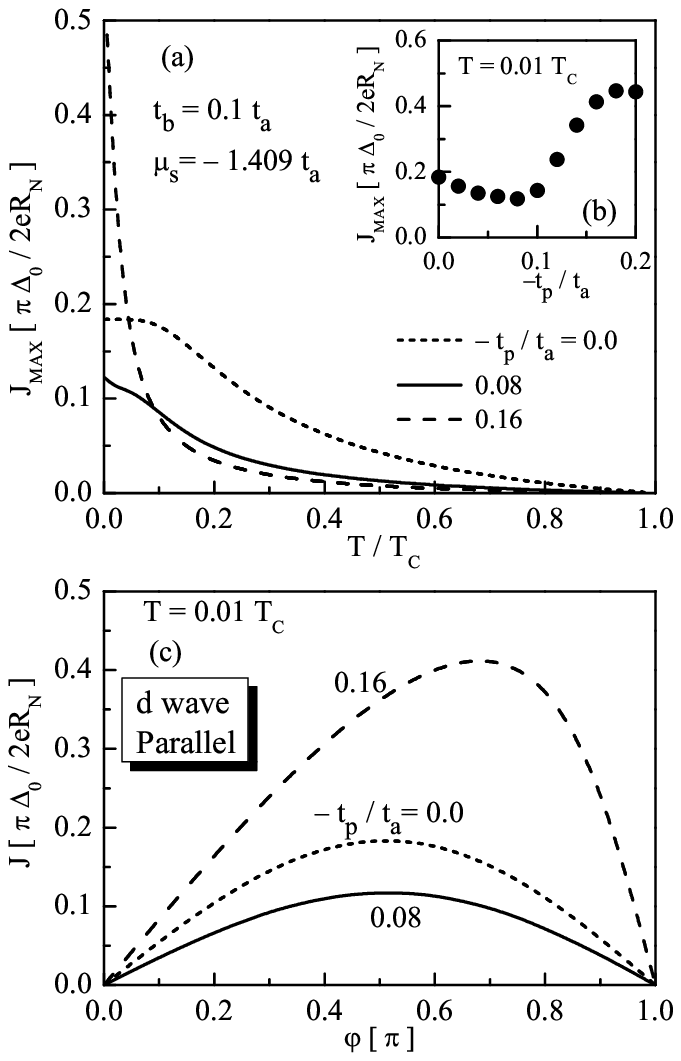}
\end{center}
\vspace{-0.5cm}
\caption{ Josephson current in the parallel junction with the $d$ wave symmetry is shown.
In (a), the Josephson current is plotted as a function of temperatures.
In (b), amplitudes of the Josephson current at $T=0.01T_c$ are shown as a function of $-t'/t_a$. 
The current-phase relation is calculated at $T=0.01T_c$ in (c).
}
\label{fig:djp}
\begin{center}
\includegraphics[width=7.0cm]{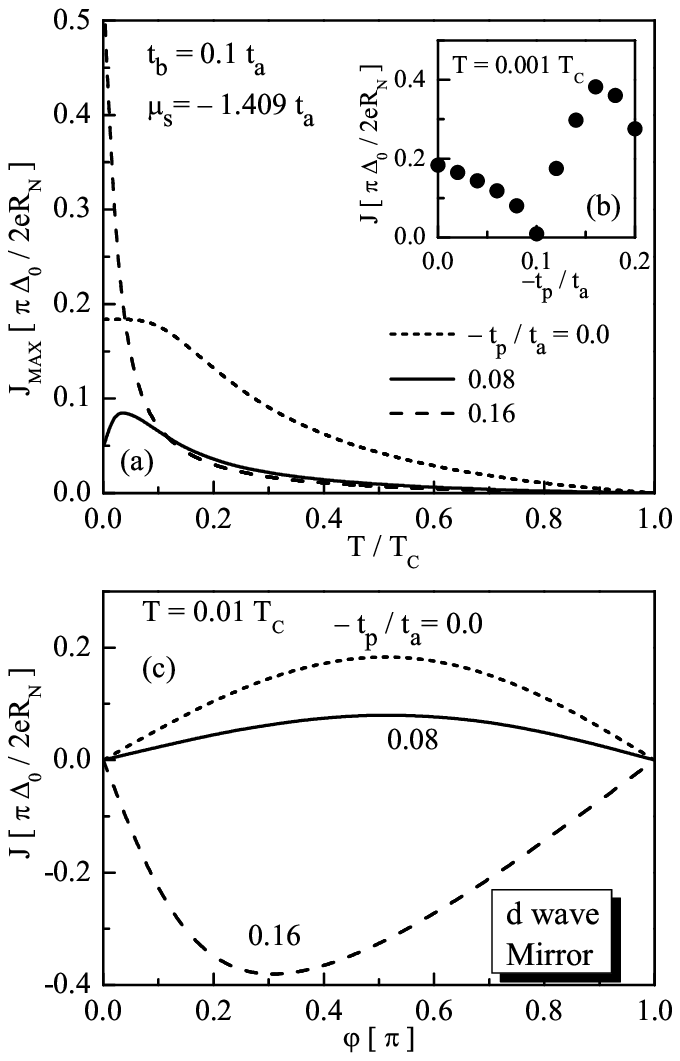}
\end{center}
\vspace{-0.5cm}
\caption{ Josephson current for the $d$ wave symmetry in the mirror-type junction.
}
\label{fig:djm}
\end{figure}

In Fig.~\ref{fig:djp}, we show the Josephson current
in the parallel junctions with the $d$ wave symmetry.  

\begin{figure}
\begin{center}
\includegraphics[width=8.0cm]{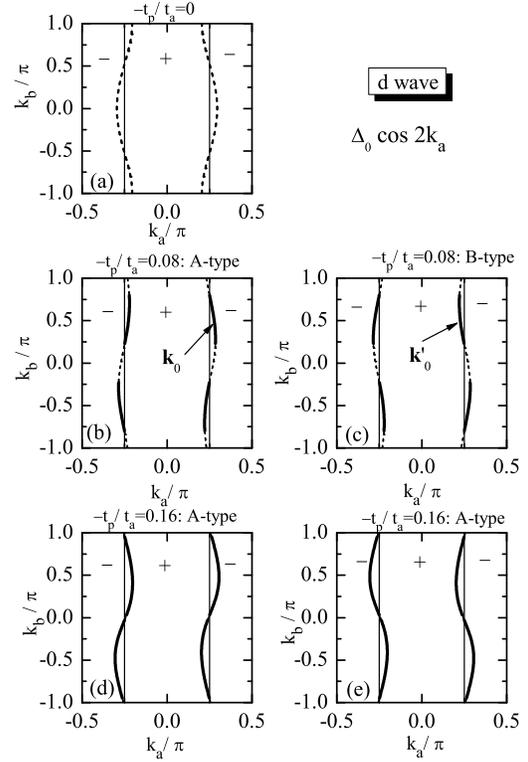}
\end{center}
\vspace{-0.5cm}
\caption{   
The Fermi surface of TMTSF for $t'=0$ is shown in (a). 
Those for $t'=-0.08t_a$ in the A-type and the B-type superconductors are shown in (b) and (c),
respectively. The pair potential in the $d$ wave symmetry is given by $\Delta \cos 2k_a$
which changes its sign at $k_a=\pm 0.25\pi$ as indicated by + and - in the figures.
The Fermi surface for $t'=-0.16t_a$ is shown for A-type and B-type superconductors in (d) and (e),
respectively.
The Fermi surface shown with the solid line satisfy the condition for the ZES in Eq.~(\ref{zes2}).
}
\label{fig:dfermi}
\end{figure}
In (a), $J_{max}$ is plotted as a function of temperatures. 
In the absence of $t'$, the Josephson current saturates in low temperatures
as that in the $s$ wave junctions. On the other hand, the results for
$-t'/t_a=$ 0.08 and 0.16 do not saturate even in low temperatures. 
In particular, the Josephson current for  $-t'/t_a=$ 0.16 rapidly increases
with decreasing temperatures. 
In the case of spin-singlet superconductors, 
a condition for appearance of the ZES is given by 
\begin{equation}
{d}(k_a,k_b) {d}(-k'_a,k_b) < 0, \label{zes2}
\end{equation}
 where ${d}(k_a,k_b)$ is the Fourier component of 
${d}_{\boldsymbol{r}-\boldsymbol{r}'}$.
In the $d$ wave symmetry, the pair potential in Eq.~(\ref{paird}) results in
\begin{equation}
d^{(d)}(k_a,k_b) = \Delta \cos( 2k_a), \label{paird2}
\end{equation}
and is the even function of $k_a$.
The additional node lines in the lattice model are $k_a =\pm 0.25 \pi$ and $\pm 0.75 \pi$.
In the absence of $t'$, Eq.~(\ref{paird2}) does not satisfy the ZEC in Eq.~(\ref{zes2}) 
for all the Fermi surface 
as shown in Fig.~\ref{fig:dfermi}, where we show the Fermi surface for $t'=0$ in (a).
The Fermi surface indicated by the broken line does not satisfy the ZEC.
In the presence of $t'$, however, the ZEC can be satisfied for some wave numbers 
on the Fermi surface as plotted with the solid line in (b) for $t'=-0.08t_a$. 
The sign of the pair potential changes frequently on the Fermi surface
because $t'$ modifies the shape of the Fermi surface lies along
the {\em additional node lines} at $k_a=\pm 0.25\pi$. 
As a result, the ZEC in Eq.~(\ref{zes2}) is satisfied for some wavenumbers, which
leads to the formation of the ZES.
This implies an importance of lattice structures on the quantum transport.
The current-phase relation deviates from the sinusoidal relation as shown in
the results for $-t'/t_a=0.16$ in Fig.~\ref{fig:djp} (c) because most of the
Fermi surface satisfy the ZEC as shown in Fig.~\ref{fig:dfermi} (d).
We note that the Fermi surface around $k_b =0$ and $\pm \pi$ are still out of
the ZEC for $-t'/t_a=0.16$.
The deviation for $-t'/t_a=0.08$ from the sinusoidal relation is smaller than 
that for $-t'/t_a=0.16$. A quasiparticle incident perpendicular to the junction 
interface mainly contributes to the Josephson current. The Fermi surface at $k_b=0$,
however, does not satisfy the ZEC as shown in Fig.~\ref{fig:dfermi} (b).
Thus the contribution of the resonant tunneling via the ZES is small for 
$-t'/t_a=0.08$ even though the temperature dependence tends to be anomalous in 
low temperatures. 

The Josephson effect in the mirror-type junctions
is different from that in the parallel junctions because of the lattice structures.
In Fig.~\ref{fig:djm}, we show the Josephson current
for the $d$ wave symmetry in the mirror-type junctions.
In (a), $J_{max}$ for $-t'/t_a$ = 0.08 first increases with decreasing
temperatures then decreases for $T<0.05T_c$. Such non monotonic
temperature dependence has been also reported in the high-$T_c$ superconductor Josephson
junctions. For $-t'/t_a$ = 0.16, the Josephson current changes its sign and the amplitude
increases rapidly with decreasing temperatures.
The differences between the parallel and the mirror-type
junctions can be understood in terms of the relative sign of the pair potentials in the
two superconductors. 
In the presence of the time-reversal symmetry, the Josephson current is decomposed 
into a series of
\begin{equation}
J=\sum_{n=1}^\infty J_n \sin(n \varphi).
\end{equation}
It was shown that the $J_1$ is roughly given by~\cite{ya01-3}
\begin{equation}
J_1 = \sum_{k_b} d_{R}(k_a',k_b) d_{L}(k_a,k_b) F_1(k_b),
\end{equation}
where $F_1$ is the positive function of $k_b$, $d_{R}(k_a',k_b)$ and $d_{L}(k_a,k_b)$
are the pair potential on the Fermi surface in the right and the left 
superconductors, respectively. 
Thus the product of $d_{R}(k_a',k_b) d_{L}(k_a,k_b)$ determines the sign of the
Josephson current proportional to $\sin\varphi$.
The parallel junction consists of two A-type superconductors as shown in Fig.~\ref{fig:system}.
Therefore $d_{R}(k_a,k_b)$ and $d_{L}(k_a,k_b)$ are identical to each other, which results in
the positive sign of $J_1$. On the other hand in the mirror-type junctions, 
$d_{L}(\boldsymbol{k}_0)$ in the A-type superconductor shown in Fig.~\ref{fig:dfermi} (b)
has the opposite sign to $d_{R}(\boldsymbol{k}'_0)$ in the B-type superconductor in (c).
It is easily shown that $d_{R}(k_a',k_b) d_{L}(k_a,k_b)$ is negative when 
the Fermi surface satisfies Eq.~(\ref{zes2}) in the mirror-type junctions.
In high temperatures, a quasiparticle on the Fermi surface shown with the broken lines in 
Figs.~\ref{fig:dfermi} (b) and (c) dominates the Josephson current
and $J_1$ is positive. This is because a quasiparticle incident perpendicular to the junction 
($k_b=0$) mainly contributes to the Josephson current and the contribution of the ZES 
is negligible.
In low temperatures, the resonant transmission via the ZES
also contributes to the Josephson current. As a consequence, $J_1$ has a non monotonic
temperature dependence as shown in Fig.~\ref{fig:djm} (a) because the 
sign of the Josephson current via the ZES is negative.
For $-t'/t_a$ = 0.16, the Josephson current is negative for low temperatures because
the resonant transmission via the ZES dominates the Josephson current.
It is also shown that the Josephson current proportional to $\sin 2\varphi$ is given 
by
\begin{equation}
J_2 = -\sum_{k_b} \left[ d_{R}(k_a,k_b) d_{L}(k_a',k_b)\right]^2 F_2(k_b),
\end{equation}
 where $F_2$ is the positive function of $k_b$.  
The sign of $J_2$ is always negative irrespective of the pairing symmetries of the
two superconductors. 
Thus the current-phase relation for $-t'/t_a$ = 0.16 in Fig.~\ref{fig:djp} (c) takes 
its maximum at $\varphi > 0.5\pi$ in the parallel junctions.
While that in the mirror-type junction takes its minimum at $\varphi < 0.5\pi$ 
in the mirror-junction in Fig.~\ref{fig:djm} (c).

\subsection{$f$ wave symmetry}
\begin{figure}
\begin{center}
\includegraphics[width=7.0cm]{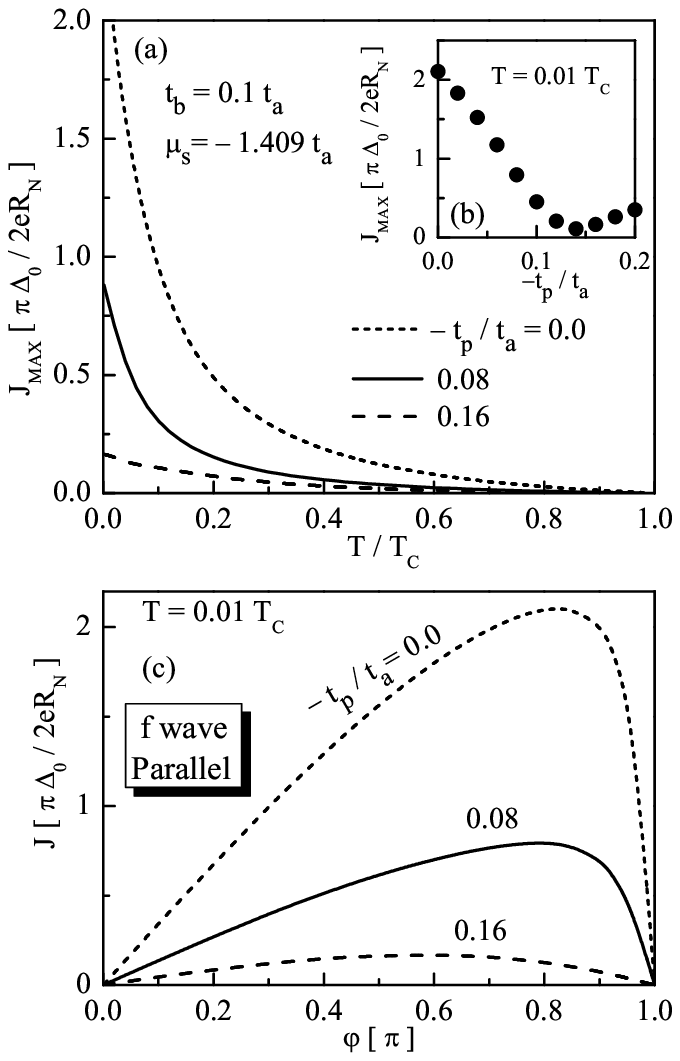}
\end{center}
\vspace{-0.5cm}
\caption{ Josephson current for $f$ wave symmetry in the parallel junctions.
In (a), the maximum amplitude of the Josephson current is plotted as a function of temperatures.
In (b), amplitudes of the Josephson current at $T=0.01T_c$ are shown as a function of $-t'/t_a$. 
The current-phase relation is calculated at $T=0.01T_c$ in (c).
}
\label{fig:fjp}
\begin{center}
\includegraphics[width=7.0cm]{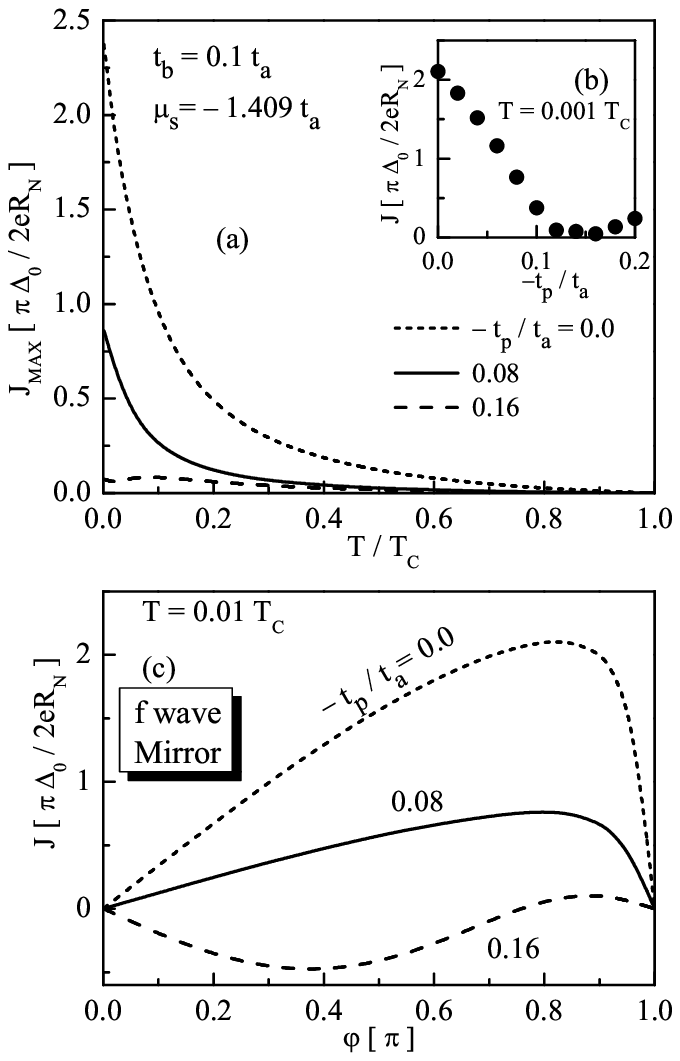}
\end{center}
\vspace{-0.5cm}
\caption{ Josephson current for $f$ wave symmetry in the mirror-type junctions.
}
\label{fig:fjm}
\end{figure}

In Fig.~\ref{fig:fjp}, we show the Josephson current
for the $f$ wave symmetry in the parallel junctions.

\begin{figure}
\begin{center}
\includegraphics[width=8.0cm]{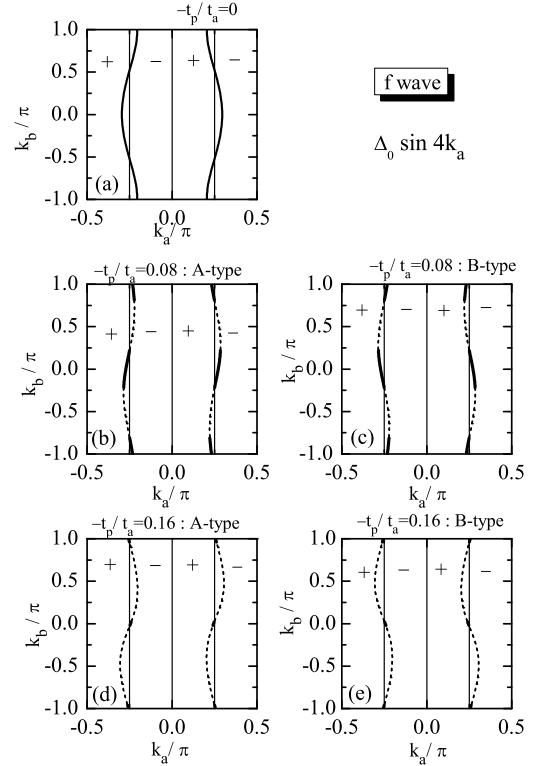}
\end{center}
\vspace{-0.5cm}
\caption{   
The Fermi surface of TMTSF for $t'=0$ is shown in (a). 
Those for $t'=-0.08t_a$ in the A-type and the B-type junctions are shown in (b) and (c),
respectively. The Fermi surface for $t'=-0.16t_a$ is shown in (d) and (e).
The pair potential in the $f$ wave symmetry is given by $\Delta \sin 4k_a$
which changes the sign at $k_a=0$ and $\pm 0.25\pi$ as indicated by + and - in the figures.
The Fermi surface shown with the solid line satisfies the condition for the ZES.
The broken line indicates the wave numbers of the Fermi surface which do not satisfy
the condition of the ZES.
}
\label{fig:ffermi}
\end{figure}
In (a), the maximum amplitude of the Josephson current is plotted as a function of temperatures 
for $-t'/t_a$ = 0, 0.08 and 0.16.
In the absence of $t'$, the Josephson current show the low-temperature anomaly
because the Fourier component of the pair potential is given by
$\Delta \sin 4k_a$ which satisfies Eq.(\ref{zes1}) for all the Fermi surface
as shown in Fig.~\ref{fig:ffermi} (a).
The additional node lines are at $k_a=\pm 0.25\pi$, $\pm 0.5\pi$ and $\pm 0.75\pi$
in the lattice model.
The anomalous behavior tends to disappear for finite $t'$. 
In Fig.~\ref{fig:ffermi} (b), the Fermi surface which do not satisfy 
Eq.(\ref{zes1}) is indicated by the broken line for $t'= -0.08t_a$.
For $t'= -0.16t_a$, most wave numbers of the Fermi surface do not
satisfy Eq.(\ref{zes1}) as shown in Fig.~\ref{fig:ffermi} (d).  
 In contrast to the $d$ wave junctions, the ZES is suppressed by 
introducing $t'$ in the $f$ wave junctions.
The phase-current relation for $-t'/t_a$ = 0 and 0.08 in Fig.~\ref{fig:fjp} (c) apparently 
deviates
from the sinusoidal function because of the multiple Andreev reflection via the ZES.
The results for $-t'/t_a$ = 0.16 slightly deviate from $\sin \varphi$, but degree of
deviation is smaller than those for $-t'/t_a$ = 0 and 0.08.

In Fig.~\ref{fig:fjm}, we show the Josephson current
for the $f$ wave symmetry in the mirror-type junctions.
In (a), the Josephson current for $-t'/t_a$ = 0 and 0.08 are essentially
the same as those in the parallel junctions. 
In the current-phase relation for $-t'/t_a=0.16$ in (c), the Josephson current becomes zero
around $\varphi=0.75\pi$. In this case, $J_1$ is negative because 
$\boldsymbol{d}_{R}(k_a',k_b) \cdot \boldsymbol{d}_{L}(k_a,k_b)$ is negative for almost 
all the Fermi surface.
The amplitude of $J_2$ is not negligible because the resonant transmission through the ZES
for a quasiparticle incident perpendicular to
the interface $(k_b=0)$ enhances $J_2$. 
In a rough estimation, we find $|J_2| \sim 0.7 |J_1|$.

\section{Discussion}

The formation of the ZES is a universal phenomenon at the surface of unconventional
superconductors~\cite{ya03-4}. It may be possible to analyze the pairing 
symmetry of unconventional superconductors from the anisotropy in the tunneling 
conductance and that in the Josephson 
current. Actually we have investigated these transport properties in both
the free electron model and the lattice model. As far as we study, 
there are no drastic differences between the two theoretical models in the case 
of high-$T_c$ superconductors.
The formation of the ZES has been believed to be insensitive to the 
electronic structures such as the shape of the Fermi surface.
The calculated results of this paper indicate that this statement is justified when 
the Fermi surface is far from the {\em additional node lines} peculiar to
the lattice structures. 
We have shown that the shape of the Fermi surface strongly affects
the formation of the ZES in Q1D organic superconductors.
The Q1D quarter-filling band and the asymmetry of the Fermi surface with 
respect to $k_a$ are important factors for the ZES to be sensitive to the 
shape of the Fermi surface.
In the quarter-filled band, the pairing interaction tends to work between 
two electrons at the second nearest neighbor sites in the $x$ direction 
as shown in Fig.~\ref{fig:system} because of the long range Coulomb repulsion.
In particular, the $f$ wave superconductivity requires the pairing correlation between
the 4th nearest neighbor sites in the $x$ direction. As a consequence, the 
pair potential changes sign at additional node lines in the first Brillouin zone such as
$k_a = \pm 0.5\pi$ and $\pm 0.25\pi$ as shown in Figs.~\ref{fig:pfermi}, ~\ref{fig:dfermi} and
~\ref{fig:ffermi}. The asymmetric hopping $(t')$ which characterizes the triangular lattice
structure causes the asymmetry of the Fermi surface with respect to $k_a$. 
The pair potential often changes its sign along the asymmetric Fermi surface because 
the Fermi surface in Q1D quarter-filled band lies along the {\em additional node
lines} at $k_a = \pm 0.25\pi$.
In the $p$ wave symmetry, characteristic behaviors of the Josephson current are 
insensitive to $t'$ because the Fermi surface is far from the {\em additional node line} 
at $k_a=\pm 0.5\pi$. 
The sign change of the pair potential around the {\em additional node lines} at 
$k_a = \pm 0.25\pi$ is very important for the formation of the ZES in the $d$ and $f$ wave
symmetries.
In the absence of $t'$, it is clear that the ZES is formed in the $f$ wave symmetry 
and that the ZES is not formed in the $d$ wave symmetry. This is because the
pair potential in the $f$ wave ($d$ wave) symmetry is an odd (even) function of $k_a$.
Roughly speaking, $t'$ assists the formation of the ZES in the $d$ wave symmetry
and suppresses the ZES in the $f$ wave symmetry.
As a results, the Josephson current exhibit the various temperature dependences and 
the phase current relations depending on the pairing symmetries, the degree of asymmetry
and the types of junctions.

\section{Conclusion}
We have studied the Josephson current in quasi one-dimensional unconventional superconductors with 
triangular lattice structures. The theoretical model describes the organic superconductors such 
as (TMTSF)$_2$X. In the calculation, we assume $p$, $d$, and $f$ wave like pairing symmetries in 
superconductors. The pair potentials have the {\em additional node lines}
because the pairing interaction works between the second or 4th nearest neighbor
sites in the current direction at the quarter-filled electron band. 
The triangular lattice structure is characterized by the asymmetric
second nearest neighbor hopping $(t')$ introduced on the square tight-binding lattice.
The zero-energy states is sensitive to $t'$ in the $d$ and $f$ wave symmetries
because the Fermi surface lies just along the additional node lines at $k_a=\pm 0.25\pi$. 
It is possible to consider the two types of junctions, (i.e.,
parallel and mirror), because of the triangular lattice structures.
The Josephson effect in parallel junctions is qualitatively different that 
in the mirror-type junctions.
We show that the Josephson current has various temperature dependences
and current-phase relations depending on the pairing symmetries of superconductors,
the shape of the Fermi surface and the types of the junctions.

%
%====Reference===================================
%
%-----------------------------------------------


\begin{thebibliography}{999}
%----------------------------------------------
%%%%%%%%%%%% TMTSF (material) %%%%%%%%%%%%%%%

\bibitem{jerome} D.~Jerome, A.~Mazaud, M.~Ribault and K.~Bechgaad: J. Phys. (France)
Lett. \textbf{41} (1980) L92. 

\bibitem{bechgaad} K.~Bechgaad, K.~Carneiro, M.~Olsen, F.~B.~Rasmussen and C.~S.~ Jacobsen: 
Phys. Rev. Lett. \textbf{46} (1981) 852. 


% Hc2 measurement
\bibitem{ijlee1} I.~J.~Lee, M.~J.~ Naughton, G.~M.~ Danner and P.~M.~Chaikin:
Phys. Rev. Lett. \textbf{78} (1997) 3555;
 I.~J.~Lee, P.~M.~Chaikin and M.~J.~Naughton: Phys. Rev. B \textbf{62} (1997) R144669.

% NMR measurement
\bibitem{ijlee2} I.~J.~Lee, S.~E.~ Brown, W.~G.~Clark, M.~J.~Strouse, M.~J.~Naughton, 
W.~Kang and P.~M.~Chaikin: Phys. Rev. Lett. \textbf{88} (2002) 017004.

% NMR measurement
\bibitem{takigawa} M.~Takigawa, H.~Yasuoka and G.~Saito: J. Phys. Soc. Jpn. \textbf{56}
 (1987) 873. 

% thermalconductivity
\bibitem{belin} S.~Belin and K.~Behnia: Phys. Rev. Lett. \textbf{79} (1997) 359.

% ZBCP
\bibitem{ha} H.~I.~Ha, J.~I.~Oh, J.~Moser and M.~J.~Naughton: Synth. Met. 
\textbf{137} (2003) 1215.


%%%%%%%%%%%%% TMTSF triplet theories %%%%%%%%%%%%%%

\bibitem{abrikosov} A.~A.~Abrikosov: J. Low. Temp. Phys. \textbf{53} (1983) 359.

\bibitem{hasegawa} Y.~Hasegawa and H.~Fukuyama: J. Phys. Soc. Jpn. \textbf{56} (1987) 877.

\bibitem{lebed} A.~G.~Lebed: Phys. Rev. B \textbf{59} (1999) R721;
A.~G.~Lebed, K.~Machida and M.~Ozaki: Phys. Rev. B \textbf{62} (2000) R795.

%%%%%%%%%%%%% TMTSF singlet theories %%%%%%%%%%%%%%

\bibitem{shimahara} H.~Shimahara: J. Phys. Soc. Jpn. \textbf{58} (1989) 1735.

\bibitem{kuroki3} K.~Kuroki and H.~Aoki: Phys. Rev. B \textbf{60} (1999) 3060.

\bibitem{kino} H.~Kino and H.~Kontani: J. Low Temp. Phys. \textbf{117} (1999) 317.


\bibitem{kuroki4} K.~Kuroki, R.~Arita and H.~Aoki: Phys. Rev. B \textbf{63} (2001) 094509.

\bibitem{pouget} J.~P.~Pouget and S.~Ravy: J. Phys. I \textbf{6} (1996) 1501.

\bibitem{kagoshima} S.~Kagoshima, Y.~Saso, M.~Maesato, R.~Kondo and T.~Hasegawa:
Solid State Commun. \textbf{110} (1999) 479.


%%%%%%%%%%%%%%% TMTSF tunnel theory %%%%%%%%%%%%%%%%%%%%%%%%%%%%%

\bibitem{tanuma1} Y.~Tanuma, Y.~Tanaka, K.~Kuroki and S.~Kashiwaya: Phys. Rev. B \textbf{64}
 (2001) 214510.

\bibitem{tanuma2} Y.~Tanuma, K.~Kuroki Y.~Tanaka, R.~Arita, S.~Kashiwaya and H.~Aoki: 
Phys. Rev. B \textbf{66} (2002) 094507. 





\bibitem{andreev} A.~F.~Andreev: Zh. Eksp. Teor. Fiz. \textbf{46} (1964) 1823.
[Sov. Phys. JETP \textbf{19} (1964) 1228].


%%%%%%%%%%% ZES %%%%%%%%%%%%%%%%%%%%%%%%%%

\bibitem{buchholtz} L.~J.~Buchholtz and G.~Zwicknagl: Phys. Rev. B \textbf{23} (1981) 5788. 

\bibitem{hu} C.~R.~Hu: Phys. Rev. Lett. \textbf{72} (1994) 1526.

\bibitem{tanaka0} Y.~Tanaka and S.~Kashiwaya: Phys. Rev. Lett. \textbf{74} (1995) 3451.

\bibitem{rpp} S.~Kashiwaya and Y.~Tanaka: Rep. Prog. Phys. \textbf{63} (2001) 1641.

\bibitem{lofwander} T.~L\"{o}fwander, V.~S.~Shumeiko and G.~Wendin:
Supercond. Sci. Technol. \textbf{14} (2001) R53.

\bibitem{ya03-4} Y.~Asano, Y.~Tanaka and S.~Kashiwaya: cond-mat/0307345.


%%%%%%%%%%%%%%%%%%%%%%%%%%%%%%%%%%%%%%%%%%%%%%%%%%%%
% d-wave in HTSC important paper
%%%%%%%%%%%%%%%%%%%%%%%%%%%%%%%%%%%%%%%%%%%%%%%%%%%%

\bibitem{tsuei} C.~C.~Tsuei and J.~R.~Kirtley: Rev. Mod. Phys. \textbf{72} (2000) 969.

\bibitem{wollman} D.~A.~Wollman, D.~J.~Van Harlingen, W.~C.~Lee, D.~M.~Ginsberg and A.~J.~Leggett:
 Phys. Rev. Lett. \textbf{71} (1993) 2134.

\bibitem{harlingen} D.~J.~van Harlingen: Rev. Mod. Phys. \textbf{67} (1995) 515. 

\bibitem{barone} V.~B.~Geshkenbein, A.~I.~Larkin and A.~Barone: Phys. Rev. B 
\textbf{36} (1987) 365. 

\bibitem{sigrist} M.~Sigrist and T.~M.~Rice: J. Phys. Soc. Jpn. \textbf{61} (1992) 4283;  
Rev. Mod. Phys. \textbf{67} (1995) 503. 

\bibitem{tanaka1994} Y.~Tanaka: Phys. Rev. Lett. \textbf{72} (1994) 3871.  


%%%%%%%%%%% ZES HTSC experiment %%%%%%%%%%%%%%%%%%%%%%%%%%

\bibitem{kashiwaya}  S.~Kashiwaya, Y.~Tanaka, M.~Koyanagi, H.~Takashima and 
K.~Kajimura: Phys. Rev. B \textbf{51} (1995) 1350.

\bibitem{kashi96} S.~Kashiwaya, Y.~Tanaka, M.~Koyanagi and K.~Kajimura:
Phys. Rev. B \textbf{53} (1996) 2667.

\bibitem{kashi95} S.~Kashiwaya, Y.~Tanaka, M.~Koyanagi and K.~Kjimura: 
J. Phys. Chem. Solids \textbf{56} (1995) 1721. 

\bibitem{alff} L.~Alff, H.~Takashima, S.~Kashiwaya, N.~Terada, H.~Ihara, Y.~Tanaka, 
M.~Koyanagi and K.~Kajimura: Phys. Rev. B \textbf{55} (1997) 14757. 

\bibitem{wang} W.~Wang, M.~Yamazaki, K.~Lee and I.~Iguchi: Phys. Rev. B \textbf{60} (1999) 4272.

\bibitem{wei} J.~Y.~T.~Wei, N.~-C.~Yeh, D.~F.~Garrigus and M.~Strasik:
 Phys. Rev. Lett. \textbf{81} (1998) 2542.

\bibitem{iguchi} I.~Iguchi, W.~Wang, M.~Yamazaki, Y.~Tanaka and S.~Kashiwaya:
Phys. Rev. B \textbf{62} (2000) R6131.  

\bibitem{geerk} J.~Geerk, X.~X.~Xi and G. Linker: Z. Phys. B \textbf{73} (1988) 329.

\bibitem{mao} Z.~Q.~Mao, M.~ M.~Rosario, K.~D.~Nelson, K.~Wu, I.~G.~Deac, 
P.~Schiffer, Y.~Liu, T.~He, K.~A.~Regan and R. J. Cava: Phys. Rev. B \textbf{67} (2003) 094502.


\bibitem{Ekin} J.~W.~Ekin, Y.~Xu, S.~Mao, T.~Venkatesan, D.~W.~Face, M.~Eddy 
and S.~A.~Wolf: Phys. Rev. B \textbf{56} (1997) 13746. 

\bibitem{Sawa1} A.~Sawa, S.~Kashiwaya, H.~Obara, H.~Yamasaki, M.~Koyanagi, Y.~Tanaka
and N.~Yoshida: Physica C \textbf{339} (2000) 107.  

\bibitem{Sawa2} H.~Kashiwaya, A.~Sawa, S.~Kashiwaya, H.~Yamazaki, M.~Koyanagi, 
I.~Kurosawa, Y.~Tanaka and I.~Iguchi:  Physica C \textbf{357}-\textbf{360} (2001) 1610.


\bibitem{Aubin} H.~Aubin, L.~H.~Greene, S.~Jian and D.~G.~Hinks:
Phys. Rev. Lett. \textbf{89} (2002) 177001.


%%%%%%%%%%%%%%%%%%%%%%%%%%%%%%%%%%%%%%%%%%%%%%%%%%%%%%
%  d-wave phase sensitive 
%%%%%%%%%%%%%%%%%%%%%%%%%%%%%%%%%%%%%%%%%%%%%%%%%%%%%%%

\bibitem{matsumoto} M.~Matsumoto and H.~Shiba: J. Phys. Soc. Jpn. \textbf{64} (1995) 1703. 

\bibitem{nagato} Y.~Nagato and K.~Nagai: Phys. Rev. B \textbf{51} (1995) 16254.   

\bibitem{ohhashi} Y.~Ohashi: J. Phys. Soc. Jpn. \textbf{65} (1996) 823; 
Y.~Ohashi and S.~Takada: J. Phys. Soc. Jpn. \textbf{65} (1996) 246.  

\bibitem{sign1} Y.~Tanuma, Y.~Tanaka, M.~Yamashiro and S.~Kashiwaya: 
Phys. Rev. B \textbf{57} (1998) 7997; 
Y.~Tanuma, K.~Kuroki Y.~Tanaka and S.~Kashiwaya: 
Phys. Rev. B \textbf{66} (2002) 174502. 


\bibitem{sign3} Y.~Tanaka, H.~Tsuchiura, Y.~Tanuma and S.~Kashiwaya: 
J. Phys. Soc. Jpn. \textbf{71} (2002) 271; 
Y.~Tanaka, Y.~Tanuma, K.~Kuroki and S.~Kashiwaya: 
J. Phys. Soc. Jpn. \textbf{71} (2002) 2102. 

\bibitem{zhu1}
H.~X.~Tang, J.-X.~Zhu and Z.~D.~Wang: Phys. Rev. B \textbf{54} (1996) 12509;
 J.-X.~Zhu, H.~X.~Tanak and Z.~D.~ Wang: Phys. Rev. B \textbf{54} (1996) 7354.  

\bibitem{stefanakis} N.~Stefanakis: Phys. Rev. B \textbf{64} (2001) 224502;
J. Phys. Cond. Matt. \textbf{13} (2001) 3643. 
  

\bibitem{wu}
S.-T.~Wu and C.-Y.~Mou: Phys. Rev. B \textbf{67} (2003) 024503;    
Phys. Rev. B \textbf{66} (2002) 012512.   

\bibitem{dong}
Z.~C.~Dong, D.~Y.~Xing and J.~Dong: Phys. Rev. B \textbf{65} (2002) 214512;    
Z.~C.~Dong, D.~Y.~Xing, Z.~D.~Wang, Z.~Zheng and J.~Dong: 
Phys. Rev. B \textbf{63} (2001) 144520.    

\bibitem{barash4}
Yu.~S.~Barash, M.~S.~Kalenkov and J.~Kurkijarvi: Phys. Rev. B \textbf{62} (2000) 6665.      

\bibitem{higashi} S.~Higashitani: J. Phys. Soc. Jpn. \textbf{66} (1997) 2556. 

%%%%%%%%%%%%%%%%%%%%%%%%%%%%%%%%%%%%%%%%%%%%%%%%%%
% d-wave  unconventional and related paper (important issue)
%%%%%%%%%%%%%%%%%%%%%%%%%%%%%%%%%%%%%%%
\bibitem{t1} Y.~Tanaka and S.~Kashiwaya: 
J. Phys. Soc. Jpn. \textbf{68} (1999) 3485;  
J. Phys. Soc. Jpn. \textbf{69} (2000) 1152.  


\bibitem{kusakabe}
Y. Tanaka, T. Hirai, K. Kusakabe and S. Kashiwaya: Phys. Rev. B \textbf{60} (1999) 6308;
T. Hirai, K. Kusakabe and Y. Tanaka: Physica C \textbf{336} (2000) 107; 
K. Kusakabe and Y. Tanaka: Physica C \textbf{367} (2002) 123; 
K. Kusakabe and Y. Tanaka: J. Phys. Chem. Solids \textbf{63} (2002) 1511.  


\bibitem{honerkamp} C.~Honerkamp and M.~Sigrist: J. Low. Temp. Phys. \textbf{111} (1998) 898;
Prog. Theor. Phys. \textbf{100} (1998) 53.

\bibitem{sengupta} K.~Sengupta, I.~\v{Z}uti\'c, H.-J.~Kwon, V.~M.~Yakovenko and S.~Das Sarma:
Phys. Rev. B \textbf{63} (2001) 144531.

\bibitem{shirai} S.~Shirai, H.~Tsuchiura, Y.~Asano, Y.~Tanaka, J.~Inoue, Y.~Tanuma 
and S.~Kashiwaya: J. Phys. Soc. Jpn. \textbf{72} (2003) 2299.

\bibitem{tsuchiura95} H.~Tsuchiura, Y.~Tanaka and Y. Ushijima:
J. Phys. Soc. Jpn \textbf{64} (1995) 922.




%%%%%%%%%%%%%% Sr2RuO4 %%%%%%%%%%%%%%%%%%%%%%%%%%%%%%%%%%%%%%
\bibitem{yama1} M.~Yamashiro, Y.~Tanaka and S.~Kashiwaya: 
Phys. Rev. B \textbf{56} (1997) 7847.

\bibitem{yama2} M.~Yamashiro, Y.~Tanaka, Y.~Tanuma and S.~Kashiwaya: 
J. Phys. Soc. Jpn. \textbf{67} (1998) 3224. 

\bibitem{yama3} M.~Yamashiro, Y.~Tanaka, N.~Yoshida and S.~Kashiwaya: 
J. Phys. Soc. Jpn. \textbf{68} (1999) 2019. 

\bibitem{barash3} Yu.~S.~Barash, A.~M.~Bobkov and M.~Fogelstrom:  
Phys. Rev. B \textbf{64} (2001) 214503.

\bibitem{ya02-2} Y.~Asano and K.~Katabuchi: J. Phys. Soc. Jpn. \textbf{71} (2002) 1974.

\bibitem{ya03-1} Y.~Asano, Y.~Tanaka, M.~Sigrist and S.~Kashiwaya: 
Phys. Rev. B \textbf{67} (2003) 184505. 

%%%%%%%%%%%%%% PrOsSb12 %%%%%%%%%%%%%%%%%%%%%%%%%%%%%%%%%%%%%%
\bibitem{ya03-3}
Y.~Asano, Y.~Tanaka, Y.~Matsuda and S.~Kashiwaya: cond-mat/0306155.


%%%%%%%%%%%%%%%%%%%%%%%%%%%%%%%%%%%%%%%%%%%%%%%%%%%%%%%%%%%5
% Ferromagnet
%
%%%%%%%%%%%%%%%%%%%%%%%%%%%%%%%%%%%%%%%%%%%%%%%%%19
\bibitem{zhu} J-X.~Zhu, B.~Friedman and C.~S.~Ting: Phys. Rev. B \textbf{59} (1999) 9558.

\bibitem{kashi2} S.~Kashiwaya, Y.~Tanaka, N.~Yoshida and M.~R.~Beasley: 
Phys. Rev. B \textbf{60} (1999) 3572.

\bibitem{zutic}
I.~Zutic and O.~T.~Valls: Phys. Rev. B \textbf{60} (1999) 6320.

\bibitem{y1} N. Yoshida, Y. Tanaka, J. Inoue and S. Kashiwaya: 
J. Phys. Soc. Jpn. \textbf{68} (1999) 1071. 

\bibitem{y2} N. Yoshida, H. Itoh, T. Hirai, Y. Tanaka, J. Inoue and S. Kashiwaya: 
Phsica C \textbf{367} (2002) 135.


\bibitem{y4} N.~Yoshida, Y.~Asano, H.~Itoh, Y.~Tanaka and J.~Inoue:
J. Phys. Soc. Jpn. \textbf{72} (2003) 895.


\bibitem{h1} T. Hirai, N. Yoshida, Y. Tanaka, J. Inoue and S. Kashiwaya: 
J. Phys. Soc. Jpn. \textbf{70} (2001) 1885. 

\bibitem{h2} T.~Hirai, Y.~Tanaka, N.~Yoshida, Y.~Asano, J.~Inoue and S.~Kashiwaya,
Phys. Rev. B \textbf{67} (2003) 174501.


%%%%%%%%%% BTRSS %%%%%%%%%%%%%%%%%%%%%%%%%%%%%%%%
\bibitem{fogelstrom} M.~Fogelstr\"{o}m, D.~Rainer and J.~A.~Sauls:
Phys. Rev. Lett. \textbf{79} (1997) 281.

\bibitem{covington} M.~Covington, M.~Aprili, E.~Paraoanu, L.~H.~Greene,
F.~Xu, J.~Zhu and C.~A.~Mirkin: Phys. Rev. Lett. \textbf{79} (1997) 277.

\bibitem{biswas} A.~Biswas, P.~Fournier, M.~M.~Qazilbash, V.~N.~Smolyaninova, H.~Balci 
and R.~L.~Greene: Phys. Rev. Lett. \textbf{88} (2002) 207004.

\bibitem{dagan} Y.~Dagan and G.~Deutscher: Phys. Rev. Lett. \textbf{87} (2001) 177004.

\bibitem{sharoni} A.~Sharoni, O.~Millo, A.~Kohen, Y.~Dagan, R.~Beck, G.~Deutscher and G.~Koren:
Phys. Rev. B \textbf{65} (2002) 134526.

\bibitem{kohen} A.~Kohen, G.~Leibovitch and G. Deutscher: Phys. Rev. Lett. \textbf{90} 
 (2003) 207005.

\bibitem{matsumoto2} M.~Matsumoto and H.~Shiba: J. Phys. Soc. Jpn. \textbf{64} (1995) 4867.

\bibitem{laughlin} R.~B.~Laughlin: Phys. Rev. Lett. \textbf{80} (1998) 5188.


\bibitem{lubimova} I.~Lubimova and G.~Koren: cond-mat/0306030.

\bibitem{kitaura} N.~Kitaura, H.~Itoh, Y.~Asano, Y.~Tanaka, J.~Inoue, Y.~Tanuma
and S.~Kashiwaya: J. Phys. Soc. Jpn. \textbf{72} (2003) 1718. 

\bibitem{TJ1}
Y.~Tanuma, Y.~Tanaka, M.~Ogata and S.~Kashiwaya: J. Phys. Soc. Jpn. \textbf{67} (1998) 1118. 

\bibitem{TJ2}
Y.~Tanuma, Y.~Tanaka, M.~Ogata and S.~Kashiwaya: Phys. Rev. B \textbf{60} (1999) 9817.

\bibitem{Tanuma2001} Y.~Tanuma, Y.~Tanaka and S.~Kashiwaya: Phys. Rev. B \textbf{64} 
 (2001) 214519.



%%%%%%% no ZBCP splitting in magnetic fields %%%%%%%%%%%%%%

\bibitem{qazilbash} M.~M.~Qazilbash, A.~Biswas, Y.~Dagan, R.~A.~Ott and R.~L.~Greene:
Phys. Rev. B \textbf{68} (2003) 024502.

\bibitem{alff2} L.~Alff, S. Kleefishch, U.~Schoop, M.~Zittartz, T.~Kemen,
T.~A.~Marx and R.~Gross: Eur. Phys. J. B \textbf{5} (1998) 423.

\bibitem{sawa3} A.~Sawa, S.~Kashiwaya, H.~Kashiwaya, H.~Obara, H.~Yamasaki, M.~Koyanagi, 
I.~Kurosawa and Y.~Tanaka: Physica C \textbf{357}-\textbf{360} (2001) 294.

\bibitem{YT022}
Y.~Tanaka, H.~Itoh, H.~Tsuchiura, Y.~Tanuma, J.~Inoue and S.~Kashiwya: 
J. Phys. Soc.  Jpn. \textbf{71} (2002) 2005. 


%%%%% quasiclassical Green Function Method d-wave interfacial roughness  %%%%%%%%% 

\bibitem{barash2} Y.~S.~Barash, A.~A.~Svidzinsky and H.~Burkhardt:
Phys. Rev. B \textbf{55} (1997) 15282.

\bibitem{golubov} A.~A.~Golubov and M.~Y.~Kupriyanov: Pis'ma Zh. Eksp. Teor. fiz 
\textbf{69} (1999) 242.[ Sov. Phys. JETP Lett. \textbf{69} (1999) 262.];
\textbf{67} (1998) 478.[ Sov. Phys. JETP Lett. \textbf{67} (1998) 501.]

\bibitem{poenicke} A.~Poenicke, Yu.~S.~Barash, C.~Bruder and V.~Istyukov:
Phys. Rev. B \textbf{59} (1999) 7102.

\bibitem{yamada} K.~Yamada, Y.~Nagato, S.~Higashitani and K.~Nagai:
J. Phys. Soc. Jpn. \textbf{65} (1996) 1540.

\bibitem{tanaka01b} Y.~Tanaka, Y.~Tanuma and S.~Kashiwaya:
Phys. Rev. B \textbf{64} (2001) 054510.

\bibitem{luck} T.~L\"{u}ck, U.~Eckern and A.~Shelankov:
Phys. Rev. B \textbf{63} (2001) 064510.

\bibitem{asai00}
Y.~Tanaka, T.~Asai, N.~Yoshida, J.~Inoue and S.~Kashiwaya:  
Phys. Rev. B  \textbf{61} (2000) R11902. 


%%%%%  d-wave interfacial roughness ZBCP splitting %%%%%%%%% 

\bibitem{ya02-1} Y.~Asano and Y.~Tanaka: Phys. Rev. B \textbf{65} (2002) 064522;
"\textit{Toward the controllable Quantum State}" Eds. H.~Takayanagi and J.~Nitta,
185, (World Scientific, Singapole, 2003). 

\bibitem{ya03-2} Y.~Asano, Y.~Tanaka and S.~Kashiwaya: cond-mat/0302287.


%\bibitem{asano96} Y.~Asano and G.~E.~W.~Bauer: Phys. Rev. B \textbf{54} (1996) 11602;
%Erratum \textbf{54} (1997) 9972.



%%%%%%%%%%%% contact with dirty normal metal %%%%%%%%%%%%%%%%%%%%%%
\bibitem{ya01-2} Y.~Asano: Phys. Rev. B \textbf{64} (2001) 014511.

\bibitem{ya02-3} Y.~Asano: J. Phys. Soc. Jpn. \textbf{71} (2002) 905; Y.~Asano: 
Physica C \textbf{367} (2002) 92; Y.~Asano: Physica C \textbf{367} (2002) 157.

\bibitem{circuit} Y.~Tanaka, Yu.~V.~Nazarov and S.~Kashiwaya:
Phys. Rev. Lett. \textbf{90} (2003) 167003.



%%%%%%%% HTSC Josephson theory %%%%%%%%%%%%%%%

\bibitem{barash} Y.~S.~Barash, H.~Burkhardt and D.~Rainer: 
Phys. Rev. Lett. \textbf{77} (1996) 4070.

\bibitem{tanaka1} Y.~Tanaka and S.~Kashiwaya: Phys. Rev. B \textbf{53} (1996) 9371.

\bibitem{tanaka2} Y.~Tanaka and S.~Kashiwaya: Phys. Rev. B \textbf{53} (1996) R11957.

\bibitem{tanaka3} Y.~Tanaka and S.~Kashiwaya: Phys. Rev. B \textbf{56} (1997) 892.

\bibitem{tanaka4} Y.~Tanaka and S.~Kashiwaya: Phys. Rev. B \textbf{58} (1998) R2948.

\bibitem{ya01-3} Y.~Asano: Phys. Rev. B \textbf{64} (2001) 224515.


\bibitem{riedel} R.~A.~Riedel and P.~F.~Bagwell: Phys. Rev. B \textbf{57} (1998) 6084. 

\bibitem{samanta} M.~P.~Samanta and S.~Datta: Phys. Rev. B \textbf{55} (1997) R8689. 


\bibitem{stefana2} N.~Stefanakis: Phys. Rev. B \textbf{65} (2002) 064533.    



%%%%%%%%%%%%%%%%%%%%%%%%%%%%%%%%%%%%%%%%%%%%%%%%%%%%%%%%%%%%%%%%%%%%%%
%  Josephson effect in cuprate  experiments
%%%%%%%%%%%%%%%%%%%%%%%%%%%%%%%%%%%%%%%%%%%%%%%%%%%%%%%%%%%%%%%%%%%%%%%%


\bibitem{Ilichev1}
E.~Il'ichev, V.~Zakosarenko, R.~P.~IJsselsteijn, H.-G.~Meyer and H.~E.~Hoenig: 
Phys. Rev. Lett. \textbf{81} (1998) 894. 


\bibitem{Ilichev2}
E.~Ilichev, M.~Grajcar, R.~Hlubina, R.~P.~IJsselsteijn, H.~E.~Hoenig, 
H.-G.~Meyer, A.~Golubov, M.~H.~S.~Amin, A.~M.~Zagoskin, A.~N.~Omelyanchouk and 
M.~Yu.~Kuprianov: Phys. Rev. Lett. \textbf{86} (2001) 53.

\bibitem{Arie}
H.~Arie, K.~Yasuda, H.~Kobayashi and I.~Iguchi: Phys. Rev. B \textbf{62} (2000) 11864. 


\bibitem{hilgenkamp1}
H.~Hilgenkamp, J.~Mannhart and B.~Mayer: Phys. Rev. B \textbf{53} (1996) 14586. 

\bibitem{tafuri}
F.~Lonbardi, F.~Tafuri, F.~Ricci, F.~Miletto Granozio, A.~Barone, G.~Testa, 
E.~Sarnelli, J.~R.~Kirtley and C.~C.~Tsuei: 
Phys. Rev. Lett. \textbf{89} (2002) 207001.  

\bibitem{smilde}
H.~J.~H.~Smilde, Ariando, D.~H.~A.~Blanko, G.~J.~Gerritsma, H.~Hilgenkamp 
and H.~Rogalla: Phys. Rev. Lett. \textbf{88} (2002) 057004. 

\bibitem{hilgenkamp2} H.~Hilgenkamp and J.~Mannhart: Rev. Mod. Phys. \textbf{74} (2002) 485. 

\bibitem{imaizumi} T.~Imaizumi, T.~Kawai, T.~Uchiyama and I.~Iguchi: 
Phys. Rev. Lett. \textbf{89} (2002) 017005.  



%%%%%%%%%%%%%%% TMTSF Tunneling theory %%%%%%%%%%%%%%%%%%%%%%%%%%%%%

\bibitem{vaccarella} C.~D.~Vaccarella, R.~D.~Duncan and C.~A.~R.~Sa de Melo:
Physica C \textbf{391} (2003) 89.


%%%%%%%%%%%%%%% TMTSF Josephson theory %%%%%%%%%%%%%%%%%%%%%%%%%%%%%
\bibitem{kwon} H.-J.~Kwon, K.~Senguputa and V.~M.~Yakovenko: cond-mat/0210148.


\bibitem{degennes} P.~G.~de~Gennes: \textit{Superconductivity of Metals 
and Alloys}, (Benjamin, New York, 1966). 





%%%%%% recursive Green function Method %%%%%%%%%%%%%%%%%  

\bibitem{lee} P.~A.~Lee and D.~S.~Fisher: Phys. Rev. Lett. \textbf{47} (1981) 882.

\bibitem{furusaki1} A.~Furusaki: Physica B. \textbf{203} (1994) 214.

\bibitem{ya01-1} Y.~Asano: Phys. Rev. B \textbf{63} (2001) 052512.

\bibitem{ya02-4} Y.~Asano: Phys. Rev. B \textbf{66} (2002) 174506.

%%%%%%%%%%%%%%%%%%%%%%%%%%%%%%%%%%%%%%%%%%%%%
%  Josephson classic biography
%%%%%%%%%%%%%%%%%%%%%%%%%%%%%%%%%%%%%%%%%%%%

\bibitem{ambegaokar}
V.~Ambegaokar and A.~Baratoff: Phys. Rev. Lett. \textbf{10} (1963) 486. 





\end{thebibliography}
\end{document}